# An Autonomous Observation and Control System Based on EPICS and RTS2 for Antarctic Telescopes

Guang-yu Zhang, Jian WANG, Peng-yi Tang, Ming-hao Jia, Jie Chen, Shu-cheng Dong, Fengxin Jiang, Wen-qing WU, Jia-jing Liu, Hong-fei Zhang

(State Key Laboratory of Technologies of Particle Detection and Electronics, Department of Modern Physics, University of Science and Technology of China, Hefei 230026, China)

Email: wangjian@ustc.edu.cn

**Abstract:** For an unattended telescopes in Antarctic, the remote operation, autonomous observation and control are essential. An EPICS (Experimental Physics and Industrial Control System) and RTS2(Remote Telescope System, 2nd Version) based autonomous observation and control system with remoted operation is introduced in this paper. EPICS is a set of Open Source software tools, libraries and applications developed collaboratively and used worldwide to create distributed soft real-time control systems for scientific instruments while RTS2 is an open source environment for control of a fully autonomous observatory. Using the advantage of EPICS and RTS2 respectively, a combined integrated software framework for autonomous observation and control is established that use RTS2 to fulfill the function of astronomical observation and use EPICS to fulfill the device control of telescope. A command and status interface for EPICS and RTS2 is designed to make the EPICS IOC (Input/Output Controller) components integrate to RTS2 directly. For the specification and requirement of control system of telescope in Antarctic, core components named Executor and Auto-focus for autonomous observation is designed and implemented with remote operation user interface based on Browser-Server mode. The whole system including the telescope is tested in Lijiang Observatory in Yunnan Province for practical observation to complete the autonomous observation and control, including telescope control, camera control, dome control, weather information acquisition with the local and remote operation.

## 1. Introduction

We build telescopes in remote sites such as high altitude plateau and mountain, Antarctic, and outer space, because they offer better seeing, less atmospheric water vapour, and reduced light pollution. But these areas are not suitable for human habitation for its harsh environments. Hence the ability of autonomous observation and control for astronomical telescope is highly required. This paper describes the design of an Antarctic autonomous observation and control system with remote operation, which is aimed at the Antarctic Bright Star Survey Telescope (BSST) which will be built in latter half of the year 2015 at Zhongshan Station, Antarctic.

BSST is a small telescope with 30cm aperture, which is used to study extrasolar planet. Its field of view is wide, so it can observe many targets simultaneously, which has advantage of bright star survey. After its construction, BSST will take advantage of polar night to make observation on wide scope sky area.

We have designed a common framework for astronomical telescopes based on the generality of control system [Jian WANG et al 2013]. The framework is modified for small telescopes in Antarctic (Fig. 1). The overall structure has changed from 5-layer to 4-layer to make it more compact and efficiency. In 4-layer structure, the Workstation Level and Local Control Level in 5-layer structure are merged into one layer – Device Control Layer. At the bottom layer is Device Layer corresponding to Hardware Layer in 5-layer structure. Observation control layer is used to control the observation of entire telescope system which correspond to Top Control Level in 5-layer structure. User layer is user interfaces for local access and remote access. Since the telescope is located in Antarctic, user needs to communicate with it through satellite channel, which is slow and expensive. The development of remote user interface needs to consider this factor.

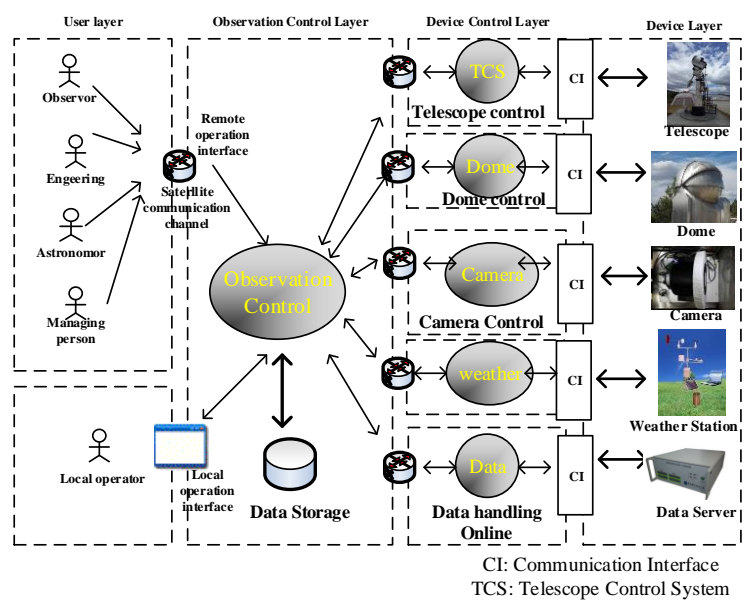

CI: Communication Interface
TCS: Telescope Control System

**Fig. 1** Structure of BSST Control System

# 2. System Structure

The structure of the autonomous observation and control system for telescope in Antarctic is designed and developed based on RTS2 and EPICS.

RTS2 (Remote Telescope System, 2nd version) [①] is a Linux based remote telescope control system created by Petr Kubanek in 2001 (Petr Kubánek et al 2006; Petr Kubánek et al 2008). It is designed for unattended operation with devices trouble-free. It can pick appropriate targets from database and make observations automatically. RTS2 is particularly suited for small-aperture telescope[②]. It has been successfully deployed on more than 10 observatories located in Spain, United States, Chile, Argentina, New Zealand, South Africa and Czech Republic.

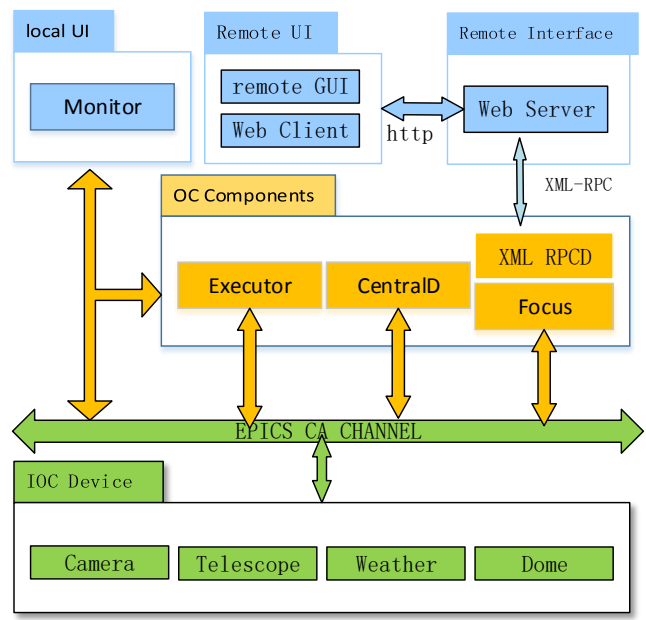

**Fig. 2** Control System Structure based on EPICS and RTS2. The components with green color correspond to Device Control Layer; the components with yellow color correspond to Observation Control Layer; the components with blue color correspond to UI Layer.

Experimental Physics and Industrial Control System (EPICS)[③] is a software environment used to develop and implement distributed soft real-time control system, which is developed by Los Alamos National Laboratory (LANL) and Argonne National Laboratory (ANL). It has been widely applied to particle accelerators, telescope and other large experiments for its real-time performance (M. Botlo, M. Jagielski, and A. Romero, 1993; Shifu Xu, Martin R. Kraimer, 2005; Jian WANG et al 2008). In the field of large telescope, an increasing number of organization choose EPICS as the basic framework of their control system, such as Keck II telescope in the United States (Lupton, W. F. 2000), Gemini 8m telescope (McGehee, P.

M. 1994; John F. Maclean, 2000), the monitor platform of Sloan Digital Sky Survey (SDSS) (McGehee, P. M. 2000).

This system is designed with a hierarchical component model. The three layers from top to bottom are UI layer, observation control layer and device control layer shown in Fig.2. The telescope control system of BSST uses EPICS to control devices, and uses RTS2 to manage observation flow, which forms a new design for telescope control and observation with the new direct interface between RTS2 connection and EPICS IOC. The underlying devices painted green are controlled by EPICS Input/Output Controller (IOC) programs which called IOC device implemented by EPICS. Observation Control(OC) components painted yellow is fulfilled using RTS2 and User Interface(UI) components painted blue is designed based on RTS2 with local operation and remote operation. This structure can reduce coupling between upper layers and bottom IOC. EPICS IOC and OC components can be adapted to each other as long as the interface stays unchanged.

UI layer contains local UI and remote UI. This layer is mainly used to provide interface for user to manage the telescope. It provides local interface and remote interface for user to access. The remote interface connects XMLRPCD (XML Remote Procedure Call Daemon) component with HTTP protocol, while the local interface connect to RTS2 components directly with socket connection. A web interface is also designed for remote operation. The functions of the web interface are as follows: statuses view of each component, image view of dome webcam, weather information view, the view and import of observation plans, autonomous observation, manual observation, common manual commands, log and history of important parameters.

The XMLRPCD in observation control layer is used to provide remote interface for remote web interface with XML-RPC protocol[1]. The protocol between OC components is the same with RTS2 protocol. The connection to bottom device control layer is Channel Access of EPICS. Some EPICS client operation needs to be added into the RTS2 Connection class, such as triggering the command Process Variable (PV), monitoring IOC status PV, so that the Connection class can connect to IOC programs.

Device control layer is composed of control programs of devices. Control programs connect to devices through standard interface such as network, serial port and USB. To make these programs being controlled by RTS2 components, some functions need to be added into these programs, such as synchronization of state and value table.

# 3. Observation control model

Programs in observation control layer are based on RTS2 components with some modifications. Connections between each observation control component are socket, while connections between observation control components and EPICS IOC are Channel Access.

---

[1] http://xmlrpc.scripting.com/default.html

## 3.1. Structure of observation control component

Centrald, XMLRPCD, Focus and Executor all belong to OC components of observation control. They are all extended from corresponding RTS2 components, so they all have similar structure. The structure of an observation control component is shown in **Fig. 3**.

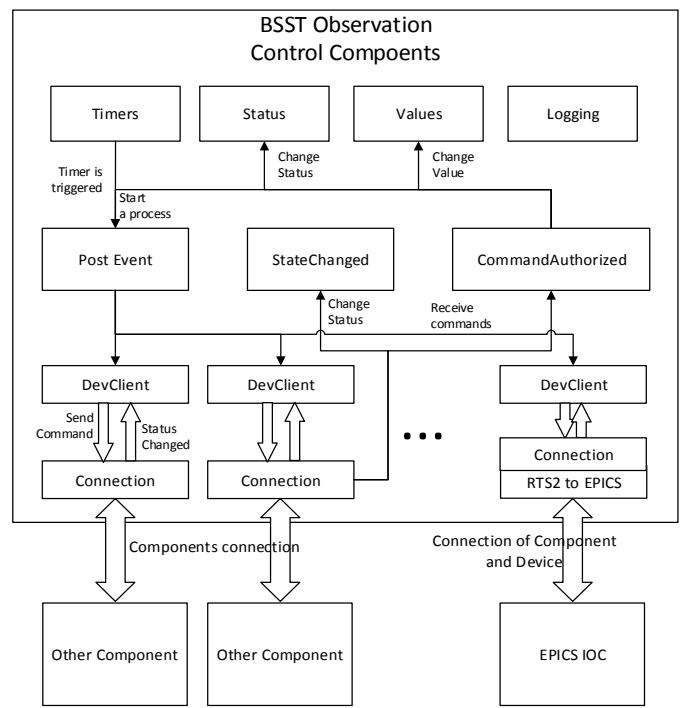

**Fig. 3** Structure of an OC component

The structure of an OC component can be divided into three levels. The first level is the functional level, including timers, values, the status value of the components, and logging function. The second level is the internal mechanism, which contains the implementation of functions and the relationship between each part. The third level is connection level, which is used to communicate with other observation components and EPICS IOCs. The connection between components is through RTS2 protocol with socket, while the connection between an OC component and an EPICS IOC is through Channel Access protocol of EPICS. The conversion between RTS2 protocol and Channel Access protocol is implemented to make it available to communicate with EPICS IOC program. The interactions of connection level fall into two categories: commands and status changes. When a command is received, connection will call a function named CommandAuthorized to process it. The CommandAuthorized function then modifies values or status, or publishes an Event through PostEvent function, depending on which command it is. BSST OC components complete a task flow through a series of Events. Each time the PostEvent function is called, a command will be sent to another component by DevClient. If the status of that component is changed, that means the Event is finished or failed. At this point, DevClient or StateChanged function will publish a new Event to continue the task.

## 3.2.  Format of communication protocol

The protocol is based on RTS2 protocol [8], string protocol with space as the separator. There are two kinds of protocol message, commands and value setting. Their formats are described as follow.

Command protocol: [command name] [param1] [param2] ...

Example: `move 12.23 45.65`

"move" is the pointing command for telescope, followed with the equatorial coordination under J2000.

Value setting protocol: X [value name] [new value]

Example: `X exposure 0.1`

This message set the exposure time of camera to 0.1 second.

Value report protocol: V [value name] [value]

Example: `V temperature 22.5`

Device may report its value when a value has changed.

Status change protocol: S [status value] [optional reason]

Example: `S 0`

This message tells other components that it has just been ready.

Command return: +/-[3 digits] [message]

When command completes, a command return is sent. Plus sign means OK; minus sign means error. The three-digit number is the return value of the command. Message contains description of error. If no error, message should be "OK".

## 3.3.  Device control layer interface

Since device control layer is based on EPICS IOC, the Connection class in OC component needs to support communication with EPICS IOC program. A middleware-like function is added into the Connection class, which is used to convert RTS2 string message to EPICS client function calls.

The conversion involves three parts: commands, values and image reading. A value in an OC component is a parameter, such as current target name, exposure time. Each RTS2 command with parameters and value is mapped to an EPICS record. Connection class uses "caput" to manipulate these records. When a command is being sent, write parameters into parameter record first, and then trigger the command record. When an image is ready, Connection will receive it and create an image data structure manually and then pass it to the image process function.

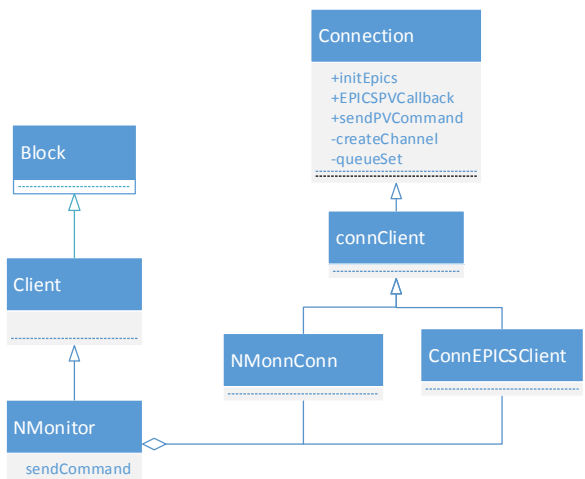

**Fig. 4** Diagram of Class Connection Extension

**Fig. 4** is the structure of Connection class and the relationship with NMonitor as an example. NMonitor is a local client of BSST that can monitor BSST status and send commands to devices. Some functions relative to EPICS communication are added into Connection class. An XML configuration file is used to represent the mapping from RTS2 commands and values to EPICS records. Connection class reads the file on initialization and creates a mapping table. Benefiting from these functions, other parts can work well with EPICS IOC without modification.

```xml
<?xml version="1.0" encoding="utf-8"?>
<EPICS>
    <CommandPV>
        <C0>
            <cmd name="expose" pv="ANDOR:cam1:Acquire">
                <arg></arg>
            </cmd>
        </C0>
    </CommandPV>
    <ValuePV>
        <C0>
            <value name="exposure"
                   pv="ANDOR:cam1:AcquireTime"
                   type="4"
                   desc="Exposure time in second"
                   fits="1"/>
        </C0>
    </ValuePV>
    <StatusPV>
        <C0>
            <status pv="ANDOR:cam1:Status" />
        </C0>
    </StatusPV>
</EPICS>
```

**Fig. 5** Format of Commands and Values Mapping file

The format of the XML configuration file is shown in **Fig. 5.** Here we take the mapping of commands as an example. Other types of mapping are similar. Command mappings are wrapped in "CommandPV" tag. Each device has a leaf tag under the "CommandPV" tag. The "cmd" tag uses attribute "name" and "pv" to represent the mapping from command name to EPICS record. In the **Fig. 5**, command "exposure" is mapped to EPICS record "ANDOR:cam1:Acquire".

A new device type "DEVICE_EPICS" is added to the RTS2 device type list to represent an EPICS IOC device. When a Connection object has been created to connect to an EPICS IOC device, it will first parse the XML configuration file and save the relationship in two mapping tables:

```
map<string(cmd),vector<string(cmdPV>>
map<string(valuePVName),string(valueName)>
```

This work is done by the "initEpics" function in Connection class. It also creates a monitor thread to monitor the changes of each EPICS PV. When a PV has changed, the monitor thread will call a callback function and pass changed message into it, and then update the corresponding value or status.

Different Connection subclasses are needed for different components or clients. The NMonitor uses NMonnConn class as its Connection class (**Fig. 4**), while RTS2 Device class use DevConnection as its Connection class. We also extend these subclasses with EPICS connection function.

To make EPICS IOCs accord with this interface, two PVs needs to be added to each IOC program to represent device name and device status. RTS2 components can check the existence of the name PV to know whether an IOC is running. The format of the device status PV should be the same with RTS2, which has been defined above. Device can have its own statuses. These statuses need to be pre-defined as macros in the "status.h" header file. The camera IOC needs another PV to show its pixel data type, the value of which needs to be consistent with RTS2.

## 3.4. Device IOC

BSST observatory system contains the following devices: telescope, camera, dome, weather station, etc. Here we take the telescope control and camera control for example.

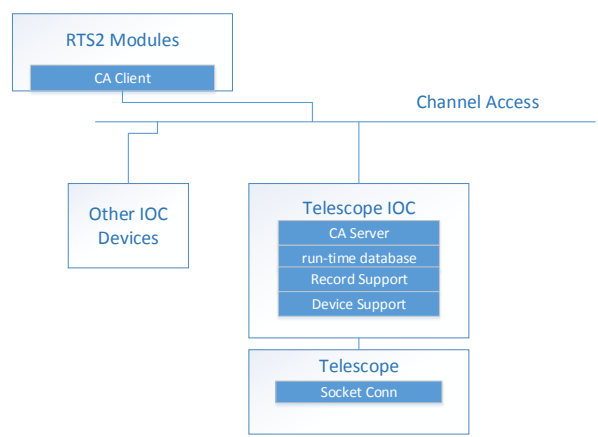

**Fig. 6** Structure of telescope IOC

The telescope in BSST is composed by mount, filter and focuser. It is controlled by specific software. The telescope IOC program is the proxy between Executor and the telescope control software. This IOC program implements the communication with telescope control software. It can convert commands sent from Executor to the message that can be recognized by the control software. The structure of telescope IOC is shown in **Fig. 6**.

Telescope IOC uses string-based socket to connect to the control software. The socket connection has two channels: command channel and alarm channel. Command channel is used to send command and return results and intermediate states, while alarm channel is used to send alarm message. Telescope IOC uses Channel Access protocol to connect to upper level components.

Image control IOC is based on a third-party software AreaDetector ⊛. Area-Detector uses camera driver to control the camera hardware. Some extra records are added to make it available to connect to RTS2, including RTS2 32-bit status variable and command interface (Guang-yu Zhang et al 2015).

# 4. Autonomous Observation

To realize autonomous observation of BSST, we investigate the need of observation and define several observation mode and observation plan. Executor can automatically load plans from database and execute them.

## 4.1. Observation modes and observation plan

Observation plan needs to be determined by objective of research. Each observation mode has a template to describe the workflow. There are mainly 5 kinds of observation modes: normal mode, dark mode, bias mode, flat mode and focus mode.

1. Normal mode: Telescope points to the target and then start tracking. Open the shutter and expose.
2. Dark mode: Close the shutter and expose for the same duration.
3. Flat mode: Shoot at dust or morning. Telescope points to a point a little higher than horizon. Telescope does not need to track.
4. Bias mode：Telescope does not need to point. Keep shutter closed and expose for 0 second.
5. Focus mode：Generally execute it every 12 hours. Choose a fairly bright star to focus the telescope.

The observation mode templates are described with XML. Commands can combined as serial or parallel sequence. These two kinds of sequence can also combine with each other to generate a tree-like command sequence.

1. "CommandSequence" is the root tag of the XML. Its "type" attribute means the observation mode. It can be set to the following macros: PLAN_STAR, PLAN_FLAT, PLAN_BIAS, PLAN_DARK and PLAN_FOCUS.
2. "SerialCommand" tag is the root of a serial command sequence. "ParallelCommand"

---

⊛ http://cars9.uchicago.edu/software/epics/areaDetectorDoc.html

tag is the root of a parallel command sequence.

3. "Command" tag expresses a command. It has some attributes to set command name, argument number, device name which it belongs to, time cost, running state and state mask. The "Arg" tag in "Command" tag expresses a parameter of the command. There can be multiple "Arg" tags in a "Command" tag. A example of "filter" command is as follow:

```
<Command name="filter" argnum="1" device="TEL" needtime="-1"
        donestate="TEL_FIL_END_MOVE" statemask="TEL_MASK_FIL">
  <Arg name="filter">$filter</Arg>
</Command>
```

A variable can defined with a "$" symbol before its name. The value of variables can be set from Executor at runtime. The "needtime" is set to "-1", which means there is no need to monitor if it is timeout.

4. "Value" tag expresses an RTS2 value. The "name" attribute is the value name in RTS2, and the "argnum" is always 1. Only one "Arg" tag is required to set the value.

```
<Value name="sensor_port_mode" argnum="1" device="CCD">
  <Arg name="port_mode">0</Arg>
</Value>
```

5. The direct child of root tag should be only one "SerialCommand" tag.

6. A "SerialCommand" tag, the same with "ParrllelCommand" tag, can have "SerialCommand", "PrallelCommand" and "Command" as its children,

7. "Command" tag can only contain "Arg" tag as its children.

## 4.2. Device Status

A device status is very important in the control system. It can reflect the current task of the device, such as pointing, tracking or switching filters for the telescope, and exposing or reading for the camera. There are also some common statuses that can be used by all devices. A change of status can be a signal of the completion of a command, which is necessary to continue an observation flow.

A device status is a 32-bit integer. The format is described as shown in Table 1.

**Table 1** Device Status Bits

| Bit1 | Bit2 | Bit3-8 | Bit9-12 | Bit13-16 | Bit17-20 | Bit21-32 |
|------|------|--------|---------|----------|----------|----------|
| Weather | Stop bit | Blocking | Weather reason | Error | Misc. | Device specific status |

Bit 1: Weather status bit. "0" means weather is good for observation; "1", defined as "BAD_WEATHER", means weather is bad. The weather status collected by weather IOC will influence this bit and the 9-12 bits described below.

Bit 2: Stop bit. When device is stopping, it will be set to 1.

Bit 3-8: Blocking bits. It is used to block other devices' actions.

Bit 9-12: Weather reason bits. It shows the specific reason of bad weather: "1" for precipitation, "2" for gale, "4" for high humidity and "8" for cloud.

Bit 13-16: Error bits. It shows if there are some errors in the device: "0" for no error, "1" for forced stopped, "2" for hardware error defined as "HW_ERROR" and "4" for not ready with some reason defined as "NOT READY". An EPICS device IOC can set its error bits according to the condition of its running condition. Executor will monitor every devices' error bits and do some process on it. The error handling in Executor is explained in section 4.3.

Bit 17-20: Miscellaneous bits such as Idle, Shutdown.

Bit 21-32: Device specific status bits. These bits are used for devices to show its specific status such as telescope is moving, dome is opening, etc.

The weather information of BSST is from a mobile weather station. The weather station provides air parameters such as temperature, relative humidity, air pressure, and wind speed. An EPICS IOC program is used to fetch these data. This program is a sensor module of BSST, and the air parameters are its Values. While it keeps reading air parameters, it will set the weather status bit and weather reason bits to notify other components.

## 4.3. Executor

Executor is a core component in BSST control system. It is used to execute observation plan, distribute commands and control the observation workflow. With a command-driven model, Executor parses observation plan to observation command sequences and parses observation commands to basic commands, and then send the commands to devices through EPICS Channel Access. It will monitor the state transition of devices to determine which command will be send. Executor can not only execute prepared observation plan, but also execute a single observation command.

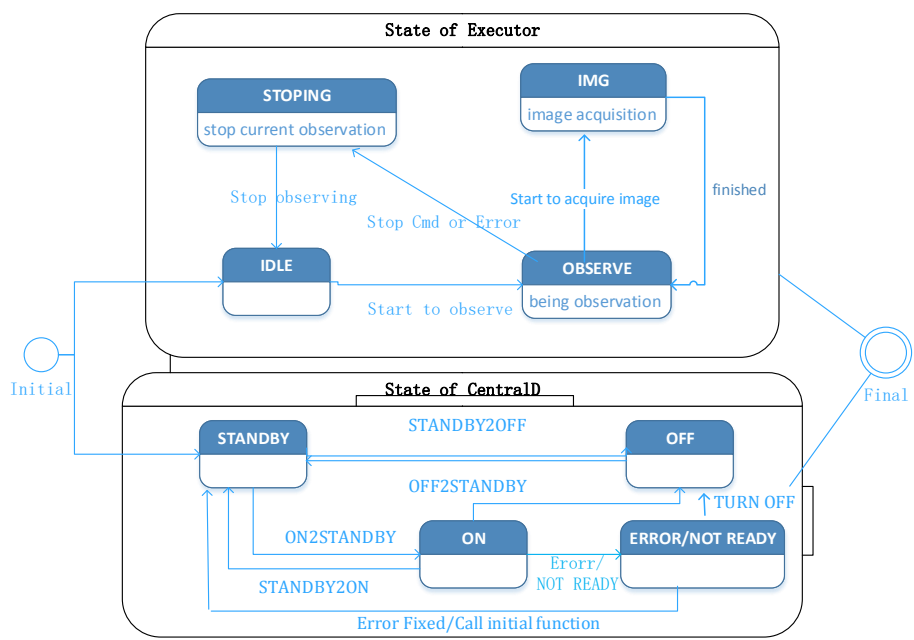

**Fig. 7** State transition diagram of Executor and CentralD

Executor can start observation automatically in automatic mode or receive a command to start observation manually. Executor needs to check if the target is available to observe (15° higher than horizon, 30° away from moon). Then Executor reads out the observation mode template from database and executes it. Device and user can interrupt the command sequence when the command sequence is running. When camera generates an image, Executor collects information needed by the FITS header and save the image to the disk. In addition, Executor needs to manage the transition of Centrald state process. Each of these main steps needs to be logged for error checking.

The key of Executor is the switching of state in different conditions. **Fig. 7** is the state transition diagram. The concrete meaning of each state is as follows:

IDLE state: Executor is idle. It can now receive commands from upper layer components and then switch to other working state.

OBSERVE state: When Executor receives an observation command in IDLE state, it will switch to OBSERVE state.

IMG state: Executor is processing image received from camera.

STOPPING state: Executor receives a stop command when observing. It will send stop command to other devices and switch state to IDLE.

Centrald has four states: ON, STANDBY, OFF and Error. The ON state means observing. The STANDBY state means it's ready for observation: camera is cooling; shutter-heating house is working; dome is opened. The OFF state means hardware is reset: camera is not cooling; telescope is parked; dome is closed; shutter-heating house is not working. When switching among these three states, Executor needs to do related work to actually switch state for other devices. If an error occurs in a device, Centrald will run into Error state and cannot be changed to other states manually, until the device is recovered from error.

There are 4 kinds of transition of Centrald state, corresponding to 4 intermediate state of Executor. ON2STANDBY means switching Centrald state from ON to STANDBY. The other three states are similar. During this state, Executor needs to stop current observation. When Executor is in STANDBY2OFF state, Executor needs to close the dome and stop camera cooling and shutter-heating house. In the other two states, Executor will do reverse actions. Executor will also check devices' error bit. If the error bit is not zero, Executor will switch to STOPPING state and then to IDLE state. Meanwhile CentralD will enter Error state. When the device is recovered from error, Centrald will change from Error state to StandBy state, waiting for observing. An alarm value is used to store error reason. See section 4.4 below for detail.

The class structure of Executor is shown in **Fig. 8**.

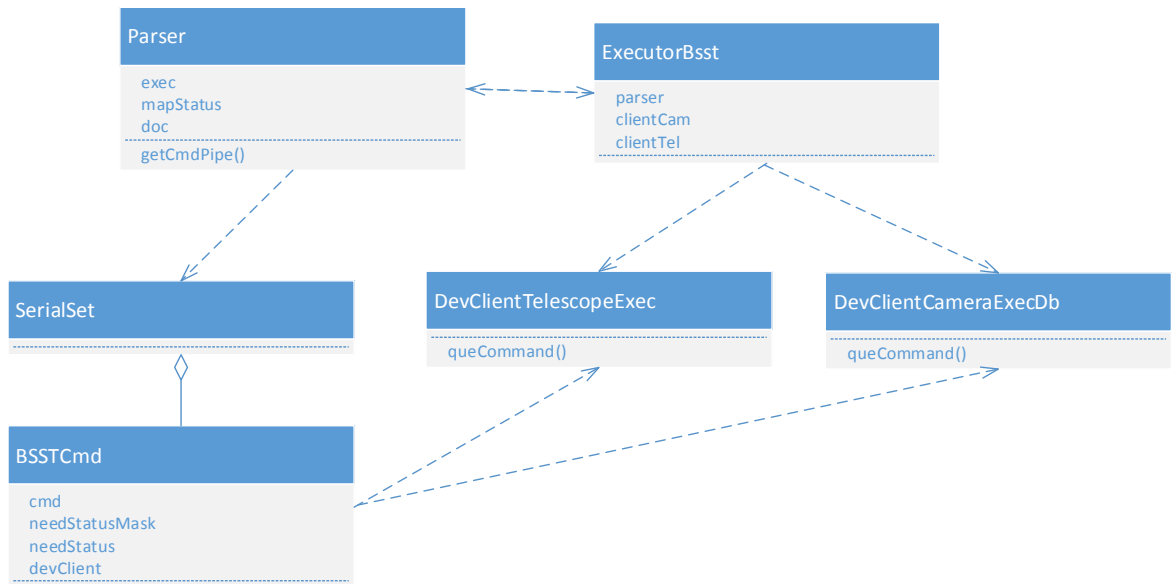

**Fig. 8** Class Diagram of Executor

When Executor receives an observation command from upper layer, it reads out the plan template and sends it to parser. The parser in Executor converts the XML format observation plan template into command sequence, and then execute it. A BSSTCmd object represents a command. It can invoke the queCommand function in DevClient to send command. DevClientTelescopeExec and DevClientCameraExecDb are two DevClients specific for telescope and camera. They can send command to devices and process the "state changed" event from devices.

Each RTS2 component has a callback function named "stateChanged". This function will be called each time when other components' state changes. The component can judge the state change and do some action. In Executor, this function is called with weather changes, command finish and device errors. When a command is finished, stateChanged will trigger the next command in current command list. The switching rule of weather state and device error state is shown in Fig. 9.

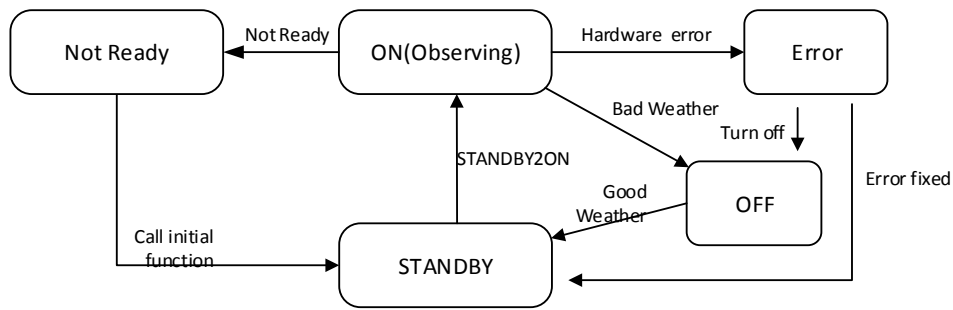

**Fig. 9** State Change of Executor

During observing, if some devices run into hardware error, telescope will switch to ERROR state. After errors are fixed, telescope will be switched to STANDBY state and finally to ON state to continue observation. When bad weather occurs, telescope will be stopped. The dome will be closed when gale or precipitation occurs. If a device is in NOT_READY state, telescope will be switched to NOT READY state which means that some device needs to be reinitialized to work properly. Executor will wait its state change and check devices' state again. If all devices are ready, Executor will switch to STANDBY state.

## 4.4. Status Monitor and Alarm Mechanism

Status Monitor and alarm mechanism is essential to the automatic control of telescope. This kind of mechanism can insure discovering error timely and reacting rapidly and thus hardware can avoid getting to damage. The system can also monitor weather change. The structure of monitor system is shown as Fig. 10.

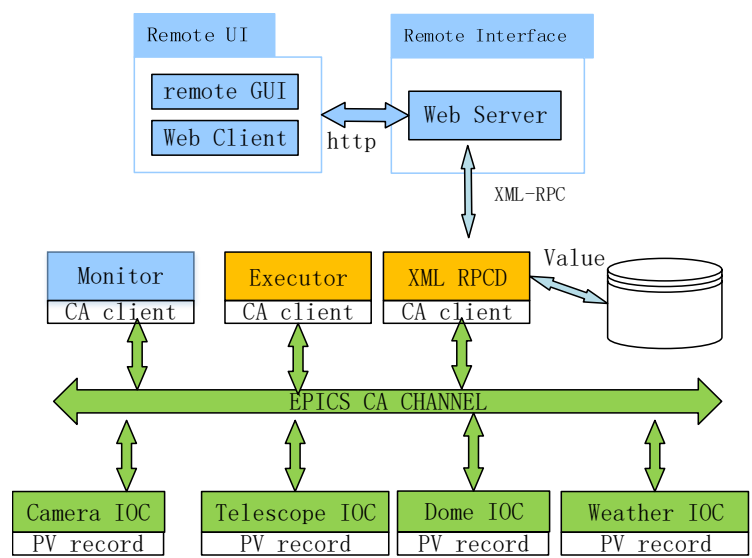

**Fig. 10** Structure of BSST Monitor System

The monitor system is consistent with the structure of BSST. There are three layers. The bottom layer is sampling layer correspond to device IOCs layer. This layer samples every status and its value, and provides uniform interface to the observation control layer. The IOCs also add alarm message into their values. The middle layer is XMLRPCD and Executor. XMLRPCD is used to collect and record values, while Executor is used to process some of the state changes. The top layer includes local monitor client and web-based remote user interface, which is used to display status and alarm message and to provide some manual operation interface.

The value recording function in XMLRPCD inherits from RTS2. XMLRPCD will read a XML configuration file to get a list of values to be recorded. When a value changes, XMLRPCD will record its new value to the database. A new API is added to get a list of value records. Web interface will draw a curve to show the trend of the value.

The alarm function is implemented with EPICS alarm status. An EPICS record can define four fields, HIGH, LOW, HIHI and LOLO to set its alarm threshold. In order to integrate EPICS alarm status with RTS2, two member variables are added into the Value class to store the alarm status. Each IOC has a record special for storing alarm message. The alarm record is a string type. Its format is shown as follows.

```
{
    "err": [ ["pvName", state, value], [...], ... ],
    "warn": [ ["pvName", state, value], [...], ... ]
}
```

The update of alarm record is done by IOCs. When an IOC gets a parameter, it will check its range and update the alarm record. Web server will keep querying each device's alarm records. If it finds an alarm record has error message, it will push the message to the web interface through WebSocket. The web interface will display a bold red number to represent the count of error messages. User can see detailed messages in the side bar.

Values to be monitored in BSST include temperature of each part, weather information and device parameters. Table 2 shows some main parameters to be monitored. As for an automatic control telescope, the more values monitored, the better they benefit the judgment and process of telescope status.

**Table 2** Threshold values of telescope parameters

| Parameter | Meaning | Warning Range | Error Range | Device State |
|-----------|---------|---------------|-------------|--------------|
| T1 | Lens Temperature 1 | $<$-70, $>$45 | $<$-80, $>$50 | HW_ERROR |
| T2 | Lens Temperature 2 | $<$-70, $>$45 | $<$-80, $>$50 | HW_ERROR |
| T5 | RA Internal | $<$-20, $>$10 | $<$-30, $>$20 | NOT_READY |
| T6 | DEC Internal | $<$-20, $>$10 | $<$-30, $>$20 | NOT_READY |
| T7 | RA Motor | $<$-20, $>$10 | $<$-30, $>$20 | NOT_READY |
| T8 | DEC Motor | $<$-20, $>$10 | $<$-30, $>$20 | NOT_READY |
| T10 | Control Case | $<$10, $>$25 | $<$5, $>$30 | HW_ERROR |
| U1 | 24V Voltage | $<$23, $>$24.5 | $<$22, $>$25 | HW_ERROR |
| U2 | 12V Voltage | $<$11.5$>$12.5 | $<$11, $>$13 | HW_ERROR |
| U3 | 5V Voltage | $<$4.7, $>$5.3 | $<$4.5, $>$5.5 | HW_ERROR |
| I1 | Mount Current | $>$2.5 | $>$3A | HW_ERROR |
| I2 | 24V Current | $>$4.5A | $>$5A | HW_ERROR |
| I3 | Heater Current of Motor and Coder | $>$2.5A | $>$3A | HW_ERROR |
| I4 | Heater Current of Motor Box | $>$2.5A | $>$3A | HW_ERROR |

| SHUTTER TEMP | Heater Temperature of Shutter | <0, >10 | <-10, >20 | NOT_READY |
|---|---|---|---|---|
| CCD_TEMP | CCD Chip Temperature | | > Target Temp | NOT_READY |
| Temperature _A | Computer Temperature | <10, >25 | <5, >30 | HW_ERROR |
| Temperature _B | MCU Temperature | <20, >50 | <10, >60 | HW_ERROR |
| WindSpeed | Wind Speed | >7m/s | >10m/s | BAD_WEATHER |
| Humidity | Humidity | >60% | >70% | BAD_WEATHER |
| Cloudy | Cloudage | >6 | >8 | BAD_WEATHER |
| Precipitation | Precipitation | | >0 | BAD_WEATHER |

The most important monitor parameters are weather information and temperatures of each part. The weather information is provided by weather station IOC, which has been described in section 4.2. Temperature information includes CCD chip temperature, shutter heater temperature, case temperature, lens barrel temperature, etc. We use the heating temperature of shutter as an example to illustrate the system. Since shutter cannot work in the environment temperature of Antarctic, a heating system is designed for it. The heating system uses a temperature sensor to sample shutter temperature. Once shutter temperature is lower than threshold, CCD will be stopped to expose. Similarly, if CCD cooling temperature does not reach the pre-set value, it will generate more noise, thus making the image data unusable. Table 2 defines the threshold of some device parameters in BSST. When an IOC program finds a device parameter is out of range, it will modify the error bit in its status. The error state will also trigger observation state change in Executor (Fig.10). When the Web server finds the temperature is out of the threshold, it will push the error message to the web interface.

Besides the parameters mentioned above, monitor information also includes disk storage information. BSST has 8 disks. The capacity of each disk is 1TB. Each two disks form a RAID1 group. Only one group is powered on at the same time in order to save energy. A disk power control board is used to switch the power of the four disk groups. When a disk group is almost full, the web interface needs to receive an alarm message. When the disk is full, an IOC program will send a switch command to the disk power control board, and the Web server will push the information to the web interface. This is the process of automatically switching disk.

## 4.5. Automatic focusing

The position of focus changes with astronomical seeing and temperature, so telescope needs to be focused before observation. The focus system is closely related to CCD control and telescope control. The structure of the focus system is shown in **Fig. 11**. When telescope has finished pointing and started to track, Executor sends focus command to BSSTFocus component. BSSTFocus chooses a position to start focusing depending on which filter is used. It first sends *focus moving* command to telescope. After that, it sends *exposure* command to camera and then analyzes the image generated by camera. When finished, BSSTFocus controls telescope to move camera again. After several times of exposure, BSSTFocus can calculate the position of focus and control telescope to move camera to that position.

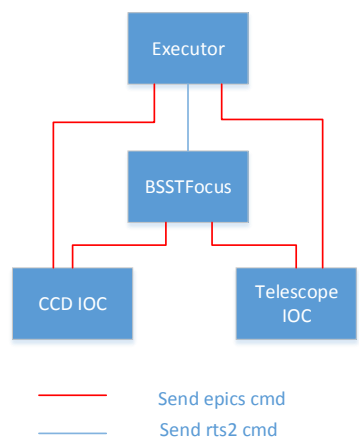
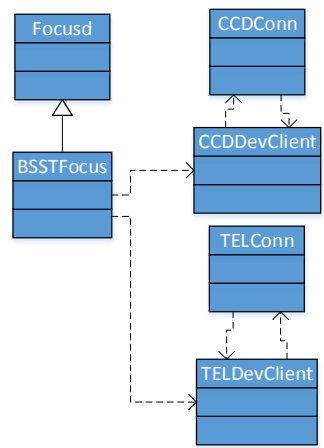

**Fig. 11** Structure of focusing system          **Fig. 12** Class Diagram of BSSTFocus

BSSTFocus is an OC component, which extends from RTS2 Focusd as shown in Fig.12. CCDDevClient and TELDevClient are DevClients to control CCD IOC and Telescope IOC. CCDConn and TelConn are the interface to interact with IOCs. The coordination among different parts is implemented with the PostEvent mechanism, which has been illustrated in **Fig. 3**.

BSSTFocus uses FWHM to evaluate the focus quality of an image. It will calculate multiple stars and take the median value as the result. It uses Sextractor® to do the calculation. The result is shown in Fig.13

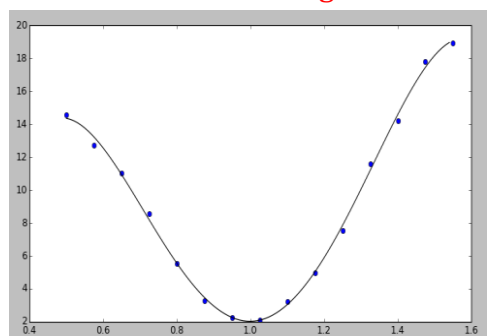

a)   Diagram of curve fitting. X-axis is imaging position and Y-axis is the mean FWHM value of stars in the image

---

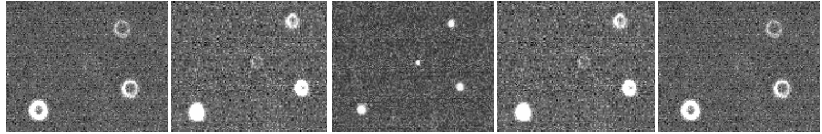

b) Sequence of stellar images near the focal point

**Fig. 13** The result of curve fitting using FWHM method

Since position of focus is related to temperature, BSSTFocus can get a mapping table from temperature to focus position using experimental data. Then it estimates the upper limit and the lower limit of the focal point. It takes several images within the limit with step of 0.1mm. The accurate focal point is calculated by making a quartic curve fitting and calculating the lowest point.

The precision of focal point calculated from curve fitting can be 0.02mm, which has reached the limitation of telescope focusing.

## 4.6. Automatic observation and remote operation

Automatic observation will start after user sends start command or the status of Centrald changes to on. After automatic observation is started, Executor tries to pick an executable plan whose start time differs from current within "waitTime". "waitTime" is a writable value in Executor. User can modify it from user interface. If an executable plan exists, Executor runs the plan. If there is no executable plan to load from database, Executor runs the last finished plan. In other case, Executor waits for "waitTime" and checks again. The "waitTime" value is aimed to insure that plans are executed according to the pre-defined time. This execution strategy is aimed to keep the efficiency of observation. The automatic observation mode will last until breaking command is received or Centrald status changes to standby or off.

The remote Web interface of BSST control system is shown in **Fig. 14**.

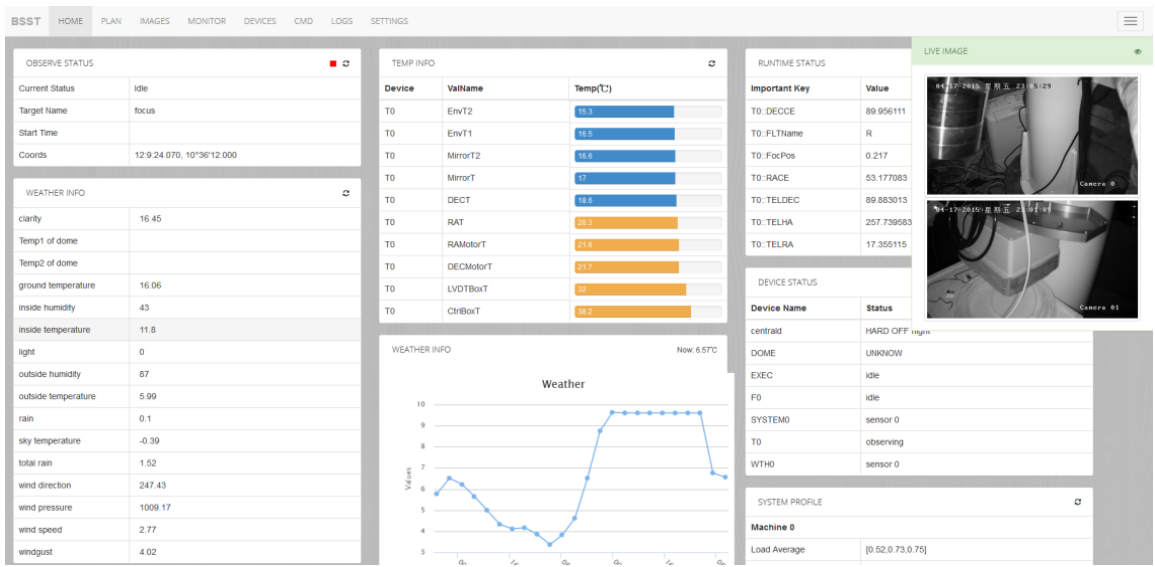

**Fig. 14** BSST remote operation interface based on web technology

The main interface includes status of every component, the image of dome webcam, weather information, view and importing of observing plan, automatic observing, manual observation and frequently-used manual commands. It also provides view of system log and history of some important parameters.

During the test of BSST in Lijiang Observatory in Yunnan Province for practical observation, it works well in the control of the entire telescope system. BSST can do 5 kinds of observing mode including normal observation. It can also finish observation task automatically. Fig.15 shows the raw image of normal observation mode.

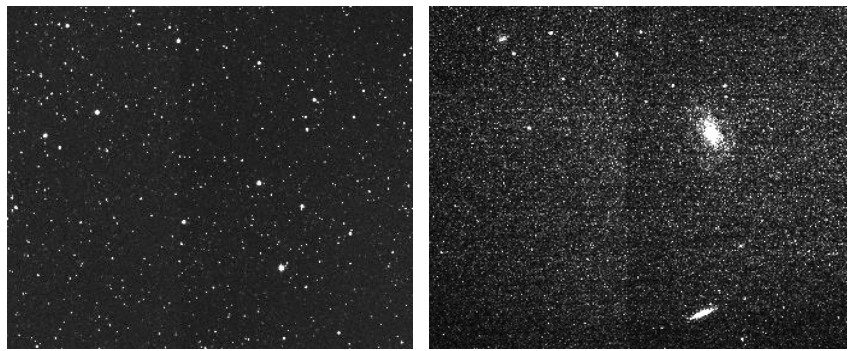

**Fig. 15** Images shot by BSST telescope

# 5. Conclusion

An autonomous observation and control system for Antarctic telescopes is designed and implemented based on RTS2 and EPICS. The system have been tested with BSST in Lijiang Observatory in Yunnan Province for practical observation. For the maturity of RTS2 and EPICS, our design is not only suitable for BSST but also for other small or mid-size telescopes in Antarctic with small modification.

# Acknowledgements

We acknowledge the financial support from BSST project, the Fundamental Research Funds for the Central Universities, National Natural Science Funds of China under Grant No: 11178020, 11275197.